\documentclass[letterpaper, 10 pt, conference]{ieeeconf}  

\IEEEoverridecommandlockouts                              
\usepackage[utf8]{inputenc}
\usepackage[T1]{fontenc}
\usepackage{graphicx}
\usepackage{amsmath}
\usepackage{authblk}
\title{\LARGE \bf
Deep Reinforcement Learning for Portfolio Optimization using Latent Feature State Space (LFSS) Module
}

\author{Kumar Yashaswi$^{1}$
\affil{Department of Mathematics, Indian Institute of Technology Kharagpur}
\thanks{*This work was supported by Department of Mathematics, Indian Institute of Technology Kharagpur}
\thanks{$^{1}$K. Yashaswi- Department of Mathematics, Indian Institute of Technology Kharagpur- kyashaswi@iitkgp.ac.in}%
}

\begin{document}

\maketitle
\thispagestyle{empty}
\pagestyle{empty}

\begin{abstract}

Dynamic Portfolio optimization is the process of distribution and rebalancing of a fund into different ﬁnancial assets such as stocks, cryptocurrencies, etc, in consecutive trading periods to maximize accumulated profits or minimize risks over a time horizon. This field saw huge developments in recent years, because of the increased computational power and increased research in sequential decision making through control theory. Recently Reinforcement Learning(RL) has been an important tool in the development of sequential and dynamic portfolio optimization theory. In this paper, we design a Deep Reinforcement Learning (DRL) framework as an autonomous portfolio optimization agent consisting of a Latent Feature State Space(LFSS) Module for filtering and feature extraction of financial data which is used as a state space for deep RL model. We develop an extensive RL agent with high efficiency and performance advantages over several benchmarks and model-free RL agents used in prior work. The noisy and non-stationarity behaviour of daily asset prices in the financial market is addressed through Kalman Filter. Autoencoders, ZoomSVD, and restricted Boltzmann machines were the models used and compared in the module to extract relevant time series features as state space. We simulate weekly data, with practical constraints and transaction costs, on a portfolio of S\&P 500 stocks. We introduce a new benchmark based on technical indicator Kd-Index and Mean-Variance Model as compared to equal weighted portfolio used in most of the prior work. The study confirms that the proposed RL portfolio agent with state space function in the form of LFSS module gives robust results with an attractive performance proﬁle over baseline RL agents and given benchmarks.
\vspace{5mm} 

Keywords- Portfolio Optimization; Reinforcement learning; Deep Learning; Kalman Filter; Autoencoders; ZoomSVD; RBM; Markowitz Model; KD-Index

\end{abstract}

\section{INTRODUCTION}

Portfolio Optimization/Management problem is to optimize the allocation of capital across various financial assets such as bonds, stocks or derivatives to optimize a preferred performance metric, like maximize expected returns or minimize risk. Dynamic portfolio optimization involves sequential decision making of continuously reallocating funds (which has roots
in control theory) into
assets in consecutive balancing periods based on real-time financial information to achieve desired performance. In financial markets, an
investor’s success heavily relies on maintaining a well balanced portfolio.
Engineering methods like signal processing [1], control theory [2], data mining [3] and advanced machine learning [4,5] are routinely used in financial market applications.
Researchers are constantly working on machine learning techniques that have proven so successful in computer vision, NLP or beating humans in chess, etc, in the domain of dynamic financial markets environment. \par
Before the advent of machine learning, most portfolio models were based on variations of Modern Portfolio Theory(MPT) by Markowitz [6]. These had the drawbacks of being static and linear in computation. The dynamic methods used for this problem like dynamic programming and convex optimization, required discrete action space based models and thus were not so efficient in capturing market information [7,8]. 
To address this issue we work with deep reinforcement learning model. Reinforcement Learning(RL) is an area in artificial intelligence which focuses on how software agents
take action in a dynamic environment to maximize cumulative performance metric or reward [9]. Reinforcement learning is appropriate in dynamical systems requiring optimal controls, like robotics [10],
self-driving cars [11] and gaming [12], with performance exceeding other models. \par
With the breakthrough of deep learning, the combination of RL and neural networks(DRL) has enabled the algorithm to give optimal results on more complex tasks and solving many constraints posed by traditional models.
In portfolio optimization, deep reinforcement learning helps in sequentially re-balancing the portfolio throughout the trading period and has continuous action space approximated by a neural network framework which circumvents the problem of discrete action space. \par
RL has been widely used in financial domain [13,14] like algorithmic trading and execution algorithms, though it has not been used to that extent for portfolio optimization. Some major works [15,16,17,18,19] gave state-of-the-art performance on their inception and most of the modern works have been inspired by these research. While [18,19] considered discrete action spaces using RL, [15,16,17] leveraged use of deep learning for continuous action space. They considered a model-free RL approach which modelled the dynamics of market through exploration. The methods proposed in [15,16,20,21,17,9] considered constraints like transaction cost and suitable reward function, though they suffered from a major drawback of state space modelling. They did not take into account the risk and dimensionality issues caused by volatile, noisy and non-stationary market environment. Asset prices are highly fluctuating and time-varying. Sudden fluctuations in asset price due to many factors like market sentiments, internal company conflict, etc cause the prices to deviate from their true value based on actual fundamentals of the asset. Many other factors like economic conditions, correlation to market, etc have an adverse effect on the prices but due to limited data and model complexity, they cannot be incorporated in modelling phase.\par
In this paper, we propose a model-free Deep RL agent with a modified state space function over the previous works which filters the fluctuations in asset price and derives a compressed time-series representation containing relevant information for modelling.
We introduce the Latent Feature State Space Module(LFSS) as a state space to our deep RL architecture.
This gives a compressed latent state space representation of filtered time series in our trading agent. It consists of 2 units:
\begin{itemize}
\item Filtering Unit 
\item Latent Feature Extractor Unit 
\end{itemize}
The filtering unit first takes the raw, unfiltered signal from asset prices and outputs a filtered signal using Kalman Filter as described in [22].
After the filtered signal is obtained the Latent Feature Extractor Unit extracts a lower dimensional latent feature space using 3 models proposed by us:
\begin{itemize}
\item Autoencoders
\item ZoomSVD
\item Restricted Boltzmann Machine
\end{itemize} 
All these 3 models are applied independently and their performance is compared to judge which is the most suitable to get a compressed latent feature space of the price signal. Autoencoders and Restricted Boltzmann Machine are self-supervised deep learning architecture that are widely used in dimensionality reduction [23,24] to learn a lower dimension representation for an input space, by training the network to ignore input noise. Svd(Singular Value Decomposition) is a matrix factorization procedure used widely in machine learning and signal processing to get a linear compressed representation of a time series. We employ ZoomSvd procedure which is a fast and memory efficient method for extracting SVD in an arbitrary query time range proposed recently in [25]. \par 
Asset price data are usually processed in OHLC(open, high, low, close) form and RL state space are built upon the same. In deep-RL models used in [15,16,20,21,9], returns or price ratio(eg High/close) are used in unprocessed form as state space with a backward time step which is the hypothesis that how much the portfolio weights are determined by the assets previous prices.
Although most of the RL agents applied in research use OHLC State Space, there are still several challenges
that LFSS Module solves: 
\begin{itemize}
\item Financial Data is considered very noisy with distorted observations due to turbulences caused by the information published daily along with those sudden market behaviours that change asset prices. These intraday noises are not the actual representative of the asset price and the real underlying state based on actual fundamentals of the asset are corrupted by noise. Filtering techniques help approximate the real state from noisy state and help's improve quality of data.
\item OHCL Data space is highly dimensional and, as such, models that try to extract relevant patterns in the raw price data can suffer from the so-called curse of dimensionality [26]. We will explore the potential of LFSS module to extract relevant features from the dynamic asset price information in a lower dimensional feature space. This latent space can hypothetically incorporate features such as economic conditions or asset fundamentals.
\end{itemize}
 
Rest of the framework is roughly similar to [1] in terms of action space, reward function, etc. We use deep deterministic policy gradients (DDPG) algorithm [27] as our RL Model with convolutional neural network(CNN) as deep learning architecture for prediction framework.

Our dataset was stocks constituting S\&P 500 with data from 2007-2015 considered for training data and 2016-2019 used for testing. 
We design a new benchmark based on technical indicator Stochastic Kd-Index and Mean-Variance model which is a variation of equal-weighted portfolio benchmark and performs much better than the same. This benchmark is based on the works in [28].

To the best of our knowledge, this is the first of many work that
leverages the use of latent features as state space, and further integrates with already known deep RL Framework in portfolio optimization domain. The main aim of our work is to investigate the effectiveness of Latent Feature State Space(LFSS) module added to our RL agent, on the process of portfolio optimization to get improved result over existing deep RL framework.

\subsection{Related Work}
With the increased complexity of deep RL models, their application in Portfolio Optimization has increased widely over the past years. As we discussed in the last section, [15,16,17,18] were one of the breakthrough papers in this domain. Our work borrowed many methods from [15,16], which had the advantage of using continuous action space. They used state of the art RL algorithms like DQN, DDPG, PPO, etc for various markets. [20] makes a comparison of model-based RL agent and model-free deep RL agent, thus showing dominance and robustness of model-free Approach. [15] showed CNN architecture performed much better than LSTM networks for the task of dynamic optimization of cryptocurrency portfolios, even though LSTM is more beneficial for time-series data. [9] used a modified reward function that prevented large portfolio weights in a single asset. \par
[26,29] were the first to use the concept of an added module for improved performance to already existing architectures. [29] added a State Augmented Reinforcement Learning(SARL) module to augment the asset information which leveraged the power of Natural Language Processing(NLP) for price movement prediction. They used financial news data to represent the external information to be encoded and augmented to the final state. [26] used combination of three modules infused prediction module(IPM), a generative adversarial data augmentation module (DAM) and a behaviour cloning module
(BCM). IPM forecast the future price movements of each asset, using historical data . DAM solves the problem of availability of large financial datasets by increasing the dataset size by making use of GANs and BCM uses a greedy strategy to reduce volatility in changes in portfolio weight's hence reducing transaction cost[26].

\section{Background}

\subsection{Portfolio Optimization}

A portfolio is a collection of multiple financial assets, and is characterized by its:
\begin{itemize}
\item Component: $n$ assets that comprise it. In our case $n=15$
\item  Portfolio vector, $w_t$: the $i^{th}$ index illustrates the proportion of the funds allocated
to the $i^{th}$ asset
     \[ w_t = [w_{1,t}, w_{2,t}, w_{3,t}, ...w_{n,t} ] \in R^n \] 
For $w_{1,t}< 0$ for any n, implies short selling is allowed.

\end{itemize}
We add a risk-free asset to our portfolio for the case if all the wealth is to be allocated to a risk-free asset. The weight vector gets modified to 
\begin{equation}
w_t = [w_{0,t}, w_{1,t}, w_{2,t}, w_{3,t}, ...w_{n,t} ]
\end{equation}
The closing price is defined as $v_{i,t}$ for asset $i$ at time $t$. The price vector $v_{t}= [v_{0,t}, v_{1,t}, v_{2,t}, v_{3,t}, ...v_{n,t} ]$, consists of the market prices of the n assets, taken as the closing price of the day in this case and constant price of risk-free asset $v_{0,t}$. Similarly, $v_{t}^{hi}$ and $v_{t}^{lo}$ denote the highest prices and the lowest price vector at time step t, respectively. 
Over time, asset price's change, therefore we denote $y_t$  as the relative price vector equal to $\frac{v_{t+1}}{v_t}$ = $(1, \frac{v_{1,t+1}}{v_{1,t}}, \frac{v_{2,t+1}}{v_{2,t}}, \frac{v_{3,t+1}}{v_{3,t}},....\frac{v_{n,t+1}}{v_{n,t}})^T$. \par
To reduce risk, portfolios with many assets are preferable over holding single assets. We assume readers have knowledge of important concepts of portfolio namely portfolio value, asset returns, portfolio variance, Sharpe ratio, mean-variance model, etc as we will not go in detail of these topics. These topics are well described in thesis reports [20,9].

\subsection{Deep RL Model as a Markov Desicion Process }

We briefly review the concepts of DRL and introduce the mathematics of the RL agent.
Reinforcement learning is a self-learning method, in which the agent interacts with the environment with no defined model and less prior information, learning from the environment by exploration while at the same time, optimally updating its strategy to maximize a performance metric. RL consist of an agent that receives the controlled state of the system and a reward associated with the last state transition. It then calculates an action which is sent back to the system. In response, the system makes a transition to a new state and the cycle is repeated as described in fig 1.
\begin{figure}[h]
\includegraphics[width=\linewidth]{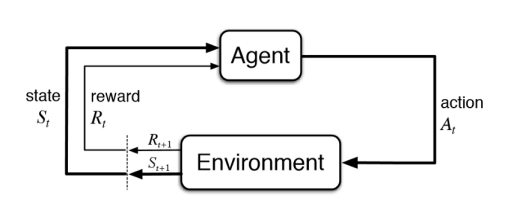}
\caption{Reinforcement Learning Setting [30]
}
\centering
\end{figure}
\par
The goal is to learn a set of actions taken for each state (policy) so as to maximize the cumulative reward in dynamic environment which in our case is the market. RL is modelled on the concepts of Markov Decision Process(MDP), a stochastic model in discrete time for decision making [9]. They work on the principle of Markov chain.\par
A MDP is defined as a 5-tuple  $(S,A,P,r,\gamma)$ described below with $t$ being time horizon over portfolio period-
\begin{itemize}
\item $S= \bigcup_t S_t$ is a collection of finite dimensional continuous state space of the RL model.
\item $A= \bigcup_t A_t$ is the finite dimensional continuous action space as we consider the market environment as
an infinite Markov Decision Process (IMDP).
\item $P: S \times A \times S \rightarrow [0,1]$ is the state transition probability function 
\item $r:  S \times A \rightarrow R$ is the instant or the expected instant reward at each time index obtained for taking an action in a particular state.
\item $\gamma \in (0,1]$ is a discount factor. When $\gamma$=1 a reward maintains its full value at each future time index independent of the relative time from present time step. As $\gamma$ decreases, the effect of reward in the future is declined exponentially by $\gamma$
\end{itemize}

\subsection{Policy Function }
Policy function($\pi$) indicate the behaviour of a RL agent in a given state. $\pi$ is a “state to action” mapping function, $\pi: S \rightarrow A$. A policy is either Deterministic $A_{t+1}=\pi(S_t)$ or Stochastic: $\pi(S|\alpha)$. We deal with the mapping to be deterministic. Since we are dealing with the case of infinite Markov decision process (IMDP), the policy is modelled as a parameterized function with respect to parameter $\theta$ and policy mapping becomes $A_t=\pi_{\theta}(S_t)$. Policy function parameterized with $\theta$ is determined solved by using DRL models like DDPG, DQN, etc which derived from concepts of Q-Learning described in the next subsection.

\subsection{Q-Learning }
Q-learning is an approach in reinforcement learning which helps in learning optimal policy function, using the concept of Q-value function. Q-value function is defined as the expected
total reward when executing action $A$ in state $S$ and henceforth follow a policy $\pi$ for future time steps. 
\begin{equation}
Q_{\pi}(S_t,A_t) = E_{\pi}(r_t | S_t,A_t)
\end{equation}
Since we are using a deterministic policy approach, we can simplify the Q-function using Bellman Equation as:
\begin{equation}
Q_{\pi}(S_t,A_t) = E_{\pi}(R_{t+1} + \gamma Q_{\pi}(S_{t+1},A_{t+1})  | S_t,A_t)
\end{equation}
The optimal policy for $Q_{\pi}$
is the one which gives maximum Q-function over all
policies:
\begin{equation}
\pi(S) = arg max Q_{\pi}(S,A)
\end{equation}
To speed up the convergence to the optimal policy, concept of replay buffer is used and a target network is used to move the relatively unstable problem
of learning the $Q$ function closer to the case of supervised learning[16].

\subsection{Deep Deterministic Policy Gradient(DDPG) }
DDPG is a model-free algorithm combining Deep Q-Network (DQN) with Deep Policy Gradient (DPG).
DQN (Deep Q-Network) stabilizes Q-function learning by replay buffer and the
frozen target network [9]. In case of DPG, the deterministic target policy function is constructed by a deep learning model and the optimal policy is reached by using gradient descent algorithm.  The action is
deterministically outputted by the policy network from the given state. DDPG works on actor-critic framework, where the actor network which outputs continuous action, and then the actor performance is improved according to critic framework which consist of an appropriate objective function. \par
The returns from the agent at a time step t is defined as $r_t= \sum_{k=t}^{\infty} \gamma^{k-t} r(S_k,A_k)$ where r is reward function, $\gamma$ is the discount factor and return $r_t$ is defined as the total discounted reward from time step $t$ to final time period[15]. 
The performance metric of $\pi_{\theta}$ for time interval $[0, t_final]$ is defined as the corresponding reward function:
\begin{equation} 
\begin{split}
J_{[0,t_{final}]}(\pi_{\theta}) = E(r, S_{\pi_{\theta}}) \\
= \sum_{t=1}^{T} \gamma^{t} r(S_t,\pi_{\theta}(S_t))
\end{split}
\end{equation}

The network is assigned random weights initially. With the gradient descent algorithm (Eq 6) weights are constantly updated to give the best expected performance metric.
\begin{equation}
\theta \leftarrow \theta + \lambda \nabla_{\theta}J_{[0,t_{final}]}(\pi_{\theta})
\end{equation}
$\lambda$ is the learning rate. For the portfolio optimization task, we modify our reward function to be similar to that described in [15] and we shall describe it in section IV. For training methodology also we borrow the works of [15] as we use the concepts of Portfolio Vector Memory(PVM) and stochastic batch optimization. We have only worked with and described in brief DDPG algorithm but other algorithms like Proximal Policy Optimization(PPO) are widely used.

\section{Latent Feature State Space Module}

Price and returns are usually derived from Open, High, Low, Close(OHLC) data. These are unprocessed and contain high level of noise. Using OHLC data based state space [15,16,20,21,9] also leads to high dimensionality and the use of extra irrelevant data. LFSS module helps in tackling such problems by using a filtering unit to reduce noise in the data and a Latent Feature Extractor Unit which helps in obtaining a set of compressed latent features which represent the asset-prices in a more suitable and relevant form. \par
The filtering unit uses Kalman Filter to obtain a linear hidden state of the returns with the noise being considered Gaussian.
The Latent Feature Extractor Unit compares 3 models Autoencoders, ZoomSVD and RBM, each being used independently with the filtering unit to extract latent feature space which is the state space of our deep RL architecture. LFSS module structure can be visualized from figure 2.

\subsection{Kalman Filter} 
Is an algorithm built on the framework of recursive Bayesian estimation, that given a series of temporal measurements containing statistical noise, outputs state estimates of hidden underlying states that tend to be more accurate representation of underlying process than those based on an observed measurement [31]. Theory of Bayesian filtering is based on hidden Markov model. Kalman Filter assumes linear modelling and gaussian error (Eq. 6,7). Kalman Filter is a class of Linear-Quadratic Gaussian Estimator(LQG) which have closed-form solution as compared to other filtering methods having approximate solutions(eg Particle Filtering)[32]. We used Kalman Filter to clean our return time-series of any Gaussian noise present due to market dynamics.
The advantages of Kalman Filter are:
\begin{itemize}
\item Gives simple but effective state estimates[18] which is computationally inexpensive
\item Market's are highly fluctuating and non-stationary, but the Kalman Filter gives optimal results despite being a LQG estimator. 
\item It is dynamic and works sequentially in real time, making predictions only using current measurement and previously estimated state on predefined model dynamics.
\end{itemize}
Though markets may not always be modelled in linear fashion and noises may not be Gaussian, Kalman filter is widely used and various experiments have proved to be effective for state estimation [31,22].
The mathematical explanation of the filter is given from eq 7-15 [22].
\begin{equation} 
x_{t+1} = F_t x_{t} + w_t,\\
\hspace{0.1cm} w_t \sim N(0,Q)
\end{equation}
\begin{equation} 
y_{t} = G_t x_{t} + v_t , \\
\hspace{0.1cm} v_t \sim N(0,R)
\end{equation}
where $x_{t}$ and $y_{t}$ are the state estimate and measurements at time index $t$ respectively. $F_{t}$ is the state transition matrix, $G_{t}$ is measurement function. White noises $w_{t}$(state noise) and $v_{t}$(process noise) are independent, normally distributed  with zero mean and constant covariance matrices Q and R respectively.

\begin{figure*}[h]
\includegraphics[width=\textwidth, height=4.5cm]{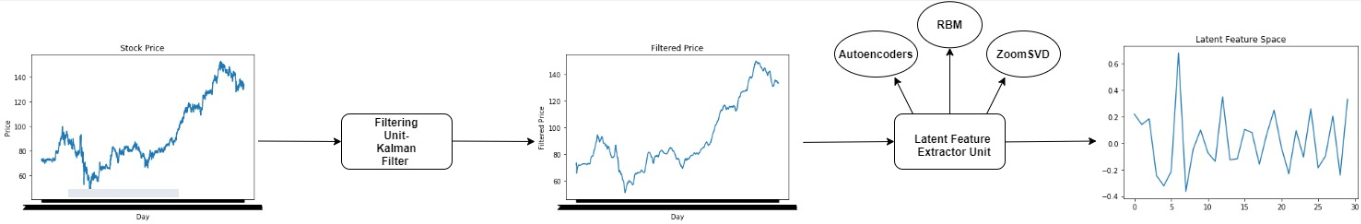}
\caption{LFSS Module Structure: (i) The stock price series is passed through a filtering layer which uses Kalman Filter to clean the signal of any Gaussian noise. (ii) The filtered signal is passed through a Latent Feature Extractor Unit which consists of one of 3 feature extractor methods a) Autoencoders b) ZoomSVD c) RBM, each applied individually and independently to obtain a lower dimensional signal
}
\centering
\end{figure*}

\begin{equation} 
x_{t+1|t+1}^{ML} = F_t x_{t|t}^{ML} + K_{t+1}\varepsilon_{t+1} 
\end{equation}
\begin{equation} 
\varepsilon_{t+1} = y_{t+1} - G_t x_{t+1|t}^{ML}
\end{equation}
\begin{equation} 
K_{t+1} = P_{t+1|t} G_t^T[G_t P_{t+1|t} G_t^T + R]^{-1}
\end{equation}
\begin{equation} 
P_{t+1|t} = F_t P_{t|t} F_t^T + Q
\end{equation}
\begin{equation} 
P_{t+1|t+1} = [I - K_{t+1}G]P_{t+1|t}
\end{equation}
\begin{equation} 
x_{0|0}^{ML} = \mu_0
\end{equation}
\begin{equation} 
P_{0|0} = P_0
\end{equation}
Eq 9-15 give the closed-form solution of Kalman filter using maximum likelihood estimation(ML). [32] gives a detailed derivation of closed-form solution using ML and Maximum a posteriori (MAP) estimate.
$\varepsilon_t$ is the error estimate defined in Eq-10 and $K_t$ is called the Kalman gain and defined in Eq-11. Eq 14-15 gives the initial state assumption of the process. In the experiment section, we use the Kalman filter for our asset price signal and input the filtered signal to the Latent Feature Extractor Unit of LFSS module.

\subsection{Autoencoders}

Its a state of the art model from non-linear feature extraction of time series leveraging the use of deep learning.
An autoencoder is a neural network architecture in which the output data is the same as input data during training, thus learns a
compact representation of the input, with no need for labels [23]. Since no output exists i.e. requires no human intervention such as data labelling it is a domain of self-supervised learning [33]. 
The architecture of an autoencoder uses two parts in this transformation[34]
\begin{itemize}
\item Encoder- by which it transforms its high dimensional inputs into a lower dimension space while
keeping the most important features
\begin{equation} 
\phi (X) \rightarrow E
\end{equation}
\item Decoder- which tries to reconstruct the original input from the output of the encoder
\begin{equation} 
\Omega (E) \rightarrow X 
\end{equation}
\end{itemize}
The output of the encoder is the latent-space representation which is of interest to us. It is a compressed form of the input data in which the most influential and significant features are kept. The key objective of autoencoders is to not directly replicate the input into the output [34] i.e learn an identity function. The output is the original input with certain information loss. Encoder should map the input to a lower dimension than input as a larger dimension may lead to learning an identity function.  \par
 The encoder-decoder weights are learnt using backpropagation similar to supervised learning approach but with output similar to the input. The mean squared norm loss is used(equation 18,19) as loss function $L$.
\begin{equation} 
\phi, \Omega= arg min_{\phi , \Omega} \|X - (\phi \circ \Omega)X \|^2
\end{equation}
\begin{equation} 
\textit{L}= \|X - X' \|^2
\end{equation}

For our methodology we compare and select the most suitable of CNN autoencoder(fig 3), LSTM autoencoders and DNN autoencoders. 
\begin{figure}[h]
\includegraphics[width=\linewidth, height=4 cm]{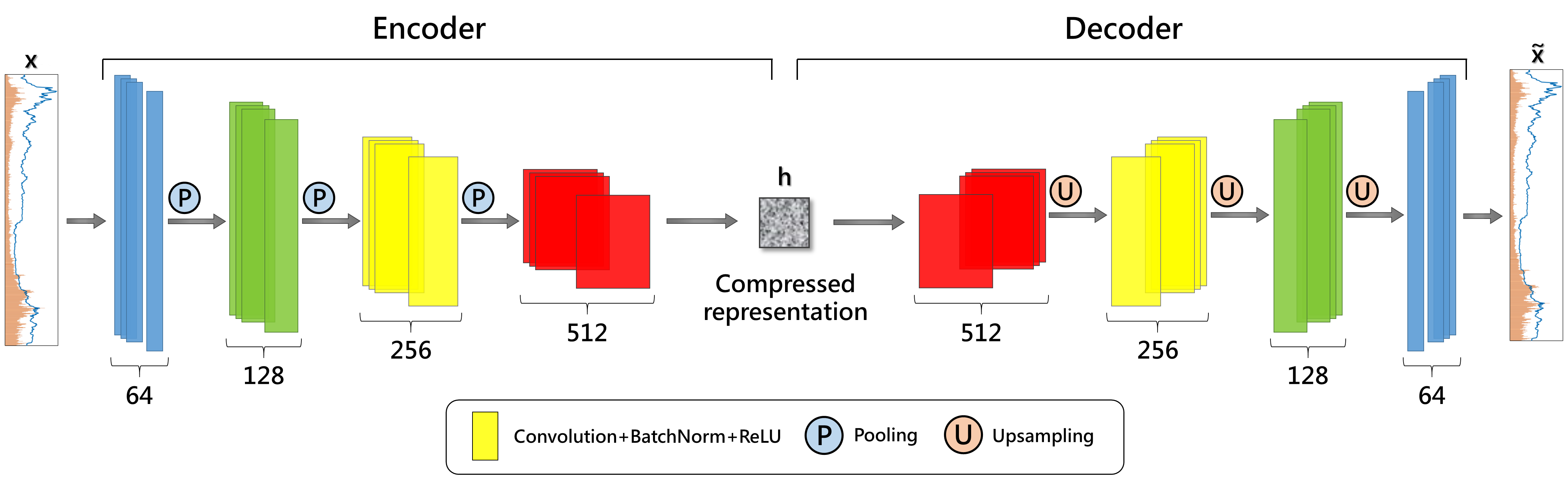}
\caption{Convolutional-Autoencoder Explained- (i) First the encoder consisting of Convolution and Pooling Layer, compresses the input time series to lower dimension. (ii) 
Decoder reconstructs the input space using upsampling and Convolution layer [35]}
\centering
\end{figure}

\subsection{ZoomSVD: Fast and Memory Efficient Method for Extracting SVD of a time series in Arbitrary Time Range-}

Based on the works in [25], this method proposes an efficient way of calculating the Singular Value Decomposition(SVD) of time series in any particular range. The SVD for a matrix is as defined in Eq 20.
\begin{figure*}[h]
\includegraphics[width=\textwidth, height=4 cm]{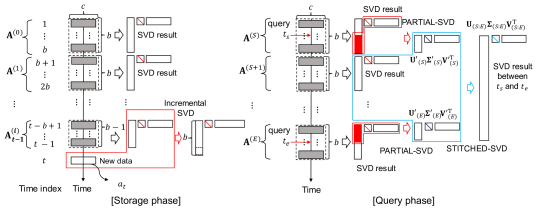}
\caption{ZoomSVD representation from [13]- (i) In the Storage-Phase the original asset-price series matrix is divided in blocks and compressed to lower dimensional SVD form using incremental-SVD. (ii) In Query-Phase, SVD in reconstructed from the compressed block structure in storage phase, for a particular time-query $[t_i, t_f]$ using Partial-SVD and Stitched-SVD [25]}
\centering
\end{figure*}
\begin{equation} 
A = U \Sigma V^T
\end{equation}
\par
If $A$ has dimensions $M \times N$(for time series case, $M$ is the length of time period and $N$ is the number of assets) then for case of compact-SVD(form of SVD used) $U$ is $M \times R$ and $V$ is $N \times R$ orthogonal matrices such that $U U^T=I_{RxR}$ and $V V^T=I_{RxR}$ and $R$ $\leq$ min$[M,N]$ is the rank of $A$. $U$ and $V$ are called left-singular vectors and right-singular vectors of $A$.  $\Sigma$=$diag(d_1,d_2,d_3,...d_R)$ is a square diagonal of size $R \times R$ with $d_1 \geq d_2 \geq d_3,...\geq d_R \geq 0$ and are called singular values of $A$. For time series data, $A = [A_1 ; A_2 ; A_3; ....A_N]$ where $A_i$ is column matrix of time series values within a range and $A$ is the vertical concatenation of each $A_i$. 
The SVD of a matrix can be calculated by many numerical methods known in linear algebra. Some of the methods are described in [36]. \par
SVD is a widely used numerical method to discover hidden/latent factors in multiple time series data [25], and in many other applications including principal component analysis, signal processing, etc. While autoencoders extract non-linear features from high dimensional time series, SVD is much simpler and finds linear patterns in the dimension of time series thus leading to less overfitting in many cases. 
Zoom-SVD incrementally and sequentially compresses time series matrix in a block by block approach to reduce the space cost in storage phase, and for a given time range query computes singular value decomposition (SVD) in query phase by methodologically stitching stored SVD results [25].
ZoomSVD is divided in 2 phases-
\begin{itemize}
\item Storage Phase of Zoom-SVD
\begin{itemize}
\item Block Matrix
\item Incremental SVD
\end{itemize}

\item Query Phase of Zoom-SVD
\begin{itemize}
\item Partial-SVD
\item Stitched-SVD
\end{itemize}
\end{itemize}
\par
Figure 4 explains the original structure proposed in [25]. In storage phase, time series matrix $A$ is divided into blocks of size $b$ decided beforehand. SVD of each block is computed sequentially and stored discarding original matrix $A$. In the query phase, an input query range [$t_i$,$t_f$] is passed and using the block structure the blocks containing the range are considered. From the selected blocks, initial and final blocks contain partial time ranges. Partial-SVD is used to compute SVD for initial and final blocks and is merged with SVD of complete blocks using Stitched-SVD to give final SVD for query range [$t_i$,$t_f$]. Further explanation of the mathematical formulation of the storage and query phase are described in [25]. \par
ZoomSVD solves the problem of expensive computational cost and large storage space in some cases. In comparison to traditional SVD methods, ZoomSVD computes time range queries up to $15$x faster and requires $15$x less space than other methods[25].
For the case of portfolio optimization problem, we use $\Sigma V^T$ as our state space. The right singular vector represents the characteristics of the time series matrix $A$ and the singular value's represents
the strength of the corresponding right singular vector [37]. U is not used because of high dimensionality due to the large time horizon required by the RL agent.

\subsection{Restricted Boltzmann machine: Probabilistic representation of the training data}
Restricted Boltzmann machine is based on the works of [38], and has similar utility to the above mentioned methods in fields like collaborative filtering, dimensionality reduction, etc.
A restricted Boltzmann machine (RBM) is a form of artificial neural network that trains to learn a probability distribution over its set of inputs and is made up of two layers: the visible and the hidden layer. The visible
layer represents the input data and the hidden layer tries to learn feature-space from the visible layer aiming to represent a probabilistic distribution of the data [39].
\begin{figure}[h]
\includegraphics[width=\linewidth, height=4 cm]{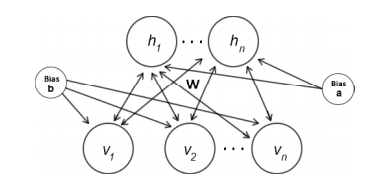}
\caption{RBM structure consisting of 3 visible units and 2 hidden units with weights w and bias a,b, forming a bipartite graph structure [39]
}
\centering
\end{figure}
Its an energy-based model implying that the probability distribution over the variables $v$(visible layer) and $h$(hidden layer) is defined by an entropy function in vector form(Eq 21).

\begin{equation} 
E(v,h) = -h^T W v - a^T v - b^T h
\end{equation}
where W is the weights, a and b are bias added. \par
RBM's have been mainly used for binary data, but there are some works [39,40]
which present new variations for dealing with continuous data using a modification to RBM called Gaussian-Bernoulli RBM
(GBRBM). We further add extension of RBM, conditional RBM(cRBM) [24,41]. The cRBM has auto-regressive weights that model short-term temporal dependencies and hidden layer unit that model long-term temporal structure in time-series data [24]. A cRBM is similar to RBM's except that the bias vector for both layers is dynamic and depends on previous visible layers [24]. We will further refer to RBM as cRBM in the paper. \par
Restricted Boltzmann machines uses contrastive divergence (CD) algorithm to train the network to maximize input probability. We refer to [39,24] for further explanation of mathematics of RBM's and cRBM's.

\section{Problem Formulation}

In the previous section, we had described RL agent formulation in terms of Markov decision process defined using 5 tuple $(S,A,P,r,\gamma)$. In this section we describe each element of tuple with respect to our portfolio optimization problem along with the dataset used and the network architecture used as our policy network.

\subsection{Market Assumptions}
In a market environment, some assumptions are considered close to reality if the trading volume of the asset in a market is high. Since we are dealing with stocks from S$\&$P 500 the trading volume is significantly high(high liquidity) to consider 2 major assumptions:
\begin{itemize}
\item Zero Slippage- Due to high liquidity, each trade can be carried out immediately at the last price when an order is
placed
\item Zero Market Impact- The wealth invested by our agent
insignificant to make an influence on the market
\end{itemize}

\subsection{Dataset}
The S$\&$P 500 is an American stock market index based on the market
capitalization of 500 large companies. There are about 500 tradable stocks constituting S$\&$P 500 index however, we will only be using a subset of 15 randomly selected stocks for our portfolio. The stocks were- Apple, Amex, Citi bank, Gilead, Berkshire Hathaway, Honeywell, Intuit, JPMC, Nike, NVDIA, Oracle, Procter $\&$ Gamble,  Walmart, Exxon Mobile and United Airlines Holdings. We took stocks from various sectors to diversify our portfolio. \par
We took the data from January 2007 to December 2019, that made up a dataset of total size of 3271 rows. 70 $\%$ (2266 prices, 2007-2015) of data was considered as training set and 30$\%$ as testing set(1005 prices, 2016-2019). This large span of time subjects our RL agent to different types of market scenarios and learns optimal strategies in both bearish and bullish market. For our risk-free asset to invest in case investing in stocks is risky, we use a bond with a constant return of 0.05 \% annually.

\subsection{Action Space}

To solve the dynamic asset allocation task, the trading agent must at every time step $t$ be able to regulate the portfolio weights $w_t$. The action $a_t$ at the time $t$ is the portfolio
vector $w_{t}$ at the time $t$:
\begin{equation} 
a_t \equiv w_{t} = [w_{1,t}, w_{2,t}, w_{3,t}, ...w_{n,t} ]
\end{equation}
The action space $A$ is thus a subset of the continuous $R^n$ real n-dimensional space:
\begin{equation} 
a_t \in A \subseteq R^n, \sum_{i=1}^{n} a_{i,t}=1, \forall t \geq 0
\end{equation}
The action space is continuous (infinite) and therefore consider the stock market as
an infinite decision-making process for MDP(IMDP) [9].
\subsection{OHLC Based State Space}
Before we describe the novel state space structure used by our RL agent, we describe the state space based on OHLC data being used in previous works. In later sections, we showcase the superior performance of our novel state space based RL agent. 

Using the data in OHLC form, [15] considered closing, high and low prices to form input tensor for state space of RL agent. The dimension of input is (n,m,3)(some works may define dimensions in (3,m,n) form), where n is the number of assets, m is the window size and 3 denotes the number of price forms. All the prices are normalized by closing price $v_{t}$ at time $t$. The individual price tensors can be written as:

\begin{equation} 
V_t = [v_{t-m+1} \oslash v_{t}| v_{t-m+2} \oslash v_{t}| v_{t-m+3} \oslash v_{t}| ..|v_{t} \oslash v_{t} ]
\end{equation}
\begin{equation} 
V_t^{(hi)} = [v_{t-m+1}^{(hi)} \oslash v_{t}| v_{t-m+2}^{(hi)} \oslash v_{t}| v_{t-m+3}^{(hi)} \oslash v_{t}| ..|v_{t}^{(hi)} \oslash v_{t} ]
\end{equation}
\begin{equation} 
V_t^{(lo)} = [v_{t-m+1}^{(lo)} \oslash v_{t}| v_{t-m+2}^{(lo)} \oslash v_{t}| v_{t-m+3}^{(lo)} \oslash v_{t}| ..|v_{t}^{(lo)} \oslash v_{t} ]
\end{equation}

Where $v_t, v_t^{(hi)} ,v_t^{(lo)}$ represents the asset close, high and low prices at time $t$, $\oslash$ represents element-wise division operator and $|$ represents horizontal concatenation. $V_{t}$ is defined individually for each asset as 
\begin{equation} 
V_t = [V_{1,t}; V_{2,t}; V_{3,t}; ..; V_{n,t} ]
\end{equation}
where $V_{i,t}$ is close price vector for asset i and $;$ represents vertical concatenation. Similar is defined for $V_t^{(hi)}$ and $V_t^{(lo)}$. $X_t$ is the final state space formed by stacking layers $V_t$, $V_t^{(hi)}$, $V_t^{(lo)}$ as seen in figure 6.
\begin{figure}[h]
\includegraphics[width=\linewidth]{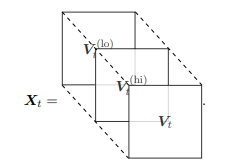}
\caption{OHLC state space in 3-dimensional input structure [15]
}
\centering
\end{figure}

In order to incorporate constraints like transaction cost, the state space $X_t$ at time t is combined with previous time step portfolio weight $w_{t-1}$ to form final state space $S_t = (X_t , w_{t-1})$. The addition of $w_{t-1}$ to state space has been widely used in other works [15,16,20,21,9] to model transaction cost of portfolio. 

\subsection{LFSS Module Based State Space}
LFSS module based state space consisted of two asset features that were processed individually through different neural networks and merged after processing. The state space considered was based on time-series features of asset returns for 15 assets in our portfolio. 
The asset feature extracted were-
\begin{itemize}
\item Covariance Matrix-  the symmetric matrix for 15 assets was calculated. This part of the state space signifies the risk/volatility and dependency part of each asset with one another. 3 matrix for close, high and low prices were computed giving a $15 \times 15 \times 3 input matrix$. 
\item LFSS Module Features- We described Filtering unit to clean the time series followed by 3 different methods Autoencoder, ZoomSVD, RBM in Latent Feature Extractor Unit to extract latent features. Close, high and low prices were individually processed by LFSS module to form 3 dimensional structure similar to figure 6 but in a lower-dimensional feature space.  
\end{itemize}

Similar to technique mentioned in last sub-section, previous time step portfolio weight $w_{t-1}$ is added to form final state space $S_t = (X_t , w_{t-1})$, where $X_t$ is merged network result.

\subsection{Reward Function}
A key challenge in portfolio optimization is controlling transaction costs(brokers’ commissions, tax, spreads, etc) which is a practical constraint to consider to not get bias in estimating returns. Whenever the portfolio is re-balanced, the corresponding transaction cost is deducted from the portfolio return. Many trading strategies work on the basic assumption on neglecting transaction cost, hence are not suitable for real-life trading and only obtain optimality using greedy algorithm(e.g allocating all the wealth into the asset which has the most promising expected growth rate) without considering transaction cost. Going by the rule of thumb the transaction commission is taken as $0.2 \%$ for every buy and sell re-balancing activity i.e $c= c_b = c_s = 0.2$. We have to consider such a reward function which takes into account practical constraints, its near-optimality with respect to portfolio performance and should be easily differentiable. Applying the reward function used in [15] satisfying all of the above conditions, the reward function at time t is :
\begin{equation} 
r_t = r(s_t,a_t) = \ln(a_t \cdot y_t -  c \sum_{i=1}^{n} |a_{i,t} - w_{i,t}|)
\end{equation}
As explained by Eq 4, the performance objective function is defined as:
\begin{equation} 
J_{[0,t_{final}]}(\pi_{\theta}) = \frac{1}{T} \ \sum_{t=1}^{T} \gamma^{t} r(S_t,\pi_{\theta}(S_t))
\end{equation}
where $\gamma$ is discount factor and $r(S_t,\pi_{\theta}(S_t))$ is the immediate reward at time t with policy function $\pi_{\theta}(S_t)$. We modify Eq 5 by dividing it by T which is used to normalize the objective function for time periods of different length. The main objective involves maximizing performance objective based on policy function $\pi_{\theta}(S_t)$ parametrized by $\theta$.
The network is assigned random weights initially. With the gradient descent algorithm(Eq 6) weights are constantly updated to give the best expected performance metric, based on the actor-critic network.
\begin{equation}
\theta \leftarrow \theta + \lambda \nabla_{\theta} J_{[0,t_{final}]}(\pi_{\theta})
\end{equation}

\subsection{Input Dimensions}
LFSS Module used considered two deep learning techniques Autoencoders and cRBM for feature extraction of time-series. While ZoomSVD has a static method for representation and we kept fixed size of cRBM structure, we compared different network topology's used for training autoencoders. 3 types of neural networks- Convolutional Neural Network(CNN), Long-Short Term Memory(LSTM) and Deep Neural Network(DNN) were compared, with the corresponding results shown in the experiments section. Our input signal consisted of considering window-size of 60 days or roughly 2 months price data, giving sufficient information at each time step $t$. The window-size is adjustable as this value is experiment based and there is no theoretical backing of choosing 60 days window. Since we are using 15 assets in our portfolio the state space concept used by [15] based on OHLC data, has input size of 15$\times$60$\times$3, where 60 is the window size, 15 is the number of assets and 3 representing 3 prices close, high and low. We added the asset covariance matrix in our state space, along with LFSS module features, which has dimensions of 15$\times$15$\times$1. The LFSS module feature space for individual methods had dimensions as follows-
\begin{itemize}
\item Autoencoder -  15$\times$30$\times$3
\item ZoomSVD - 15$\times$15$\times$3
\item cRBM - 15$\times$30$\times$3
\end{itemize}
Autoencoders and cRBM provide flexibility of reducing the original price input into variable-sized feature space, while ZoomSVD has drawback of fixed-sized mapping.

\subsection{Network Architecture}
\begin{figure*}[h]
\includegraphics[width=\textwidth, height=5 cm]{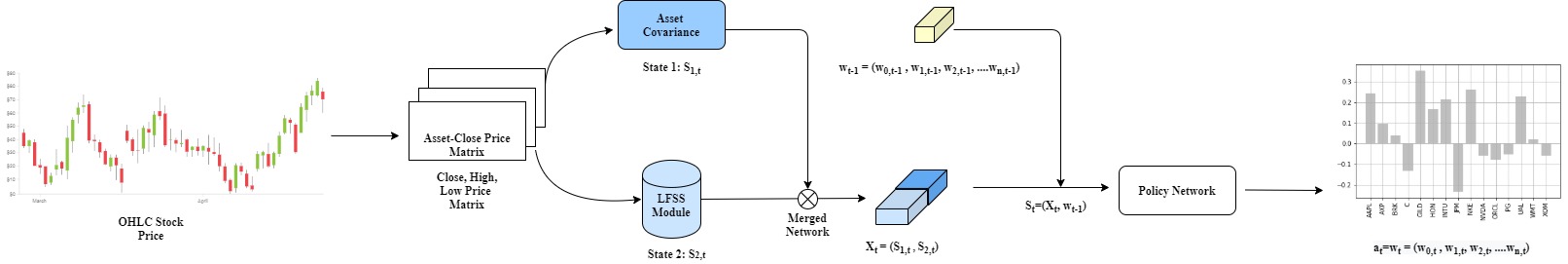}
\caption{The framework of our proposed LFSS Module based RL Agent. Asset prices are structured in 3-dimensional close, high and low price and then the state space $S_{1,t}$ and $S_{2,t}$ are extracted from state-covariance and  LFSS module, and a combined state $X_{t}$ is formed from a merged Neural Network. Combined with previous timestep action space $a_{t-1}$, a new portfolio vector is formed for time t through the policy network}
\centering
\end{figure*}
Our Deterministic target policy function is constructed by a neural network architecture used by the agent. Several variations of architecture are used for building policy function revolving around CNN, LSTM, etc. To establish the performance of Latent Feature State Space compared to OHLC state space, we work with single model architecture for better judgement of performance of various state-spaces. Considering the results in [15,20,21], CNN based network outperformed RNN and LSTM for approximating policy function, due to its ease in handling large amounts of multidimensional data, though this result is empirical as sequential neural network like RNN and LSTM should model price data better than CNN[15]. \par
Motivated by [15] we used similar techniques, namely Identical Independent Evaluators(IIE) and Portfolio-Vector Memory(PVM) for our trading agent. In IIE, the policy network propagates independently for the n+1 assets with common network parameters shared between each flow. The scalar output of each identical stream is normalized by the softmax layer and compressed into a weight vector as the next period’s action. PVM allows simultaneous mini-batch training of the network, enormously improving training efficiency [15]. 
\par
Our state space consists of 3 parts- LFSS Module Features, covariance matrix and the portfolio vector from the previous rebalancing step($w_{t-1}$). LFSS and covariance matrix were individually processed through CNN architecture, and concatenated together into a single vector $X_t$ after individual layers of processing. The previous actions, $w_{t-1}$, is concatenated to this vector to form final stage state $S_t= (X_t, w_{t-1})$. This is passed through a deep neural network, with IIE and PVM setup, to output portfolio vector $w_t$ with the training process as described in the previous subsection. The basic layout of our RL agent is as given in figure 7.  

\subsection{Benchmarks}
\begin{itemize}
\item Equal Weight Baseline(EW) - is a trivial baseline which assigns equal weight to all assets in the portfolio. 
\item S\&P 500 - the stock-market index which depicts the macro-level movement of the market as a whole.
\item OHLC State-Space Rl Agen(Baseline-RL) - uses unprocessed asset prices as state space and trains the agent with a CNN approximated policy function. It is the most widely used RL agent structure for dynamic portfolio optimization.
\item  Weighted Moving Average Mean Reversion(WMAMR) - is a method which
captures information from past period price relatives using moving averaged loss function to achieve an optimal portfolio.
\item Integrated Mean-Variance Kd-Index Baseline(IMVK)- derived from [28], this works on selecting a subset of given portfolio asset based on technical-indicator Kd-Index and mean-variance model[6], and allocates equal-weights to the subset at each time-step. For calculating Kd-Index first a raw stochastic value(RSV) is calculated as follows:
\begin{equation} 
RSV_t = \frac{(v_t - v_{min})}{(v_{max}- v_{min})} \times 100
\end{equation}
$v_t, v_{max}, v_{min}$ represent the closing price at time t, highest closing price and lowest closing price, respectively. Kd-Index is computed as:
\begin{equation} 
K_t = RSV_t \times \frac{1}{3} + K_{t-1} \times \frac{2}{3}
\end{equation}
\begin{equation} 
D_t = K_t \times \frac{1}{3} + D_{t-1} \times \frac{2}{3}
\end{equation}
where $K_1 = D_1 =50$. Whenever $K_{t-1} \leq D_{t-1}$ and  $K_{t} \textgreater D_{t}$, a buy signal is created.
Two strategies, Moderate and aggressive, are formed upon Kd-Index. Moderate strategy works on selecting an asset only if Mean-Variance model weights for the asset is positive and Kd-Index indicated a buy signal at that time-step. Aggressive strategy selects an asset only based on Kd-Index indicating a buy signal. Equal distribution of wealth is done on the subset.
\end{itemize}

\section{Experimental Results}

\subsection{Filtering Unit}
As described before, LFSS Module has an added Filtering Unit consisting of a Kalman Filter to filter out noise from price data of each asset. Since deep learning model are sensitive to the quality of data used, many complex deep learning models may overfit the data, learning noise in the data as well, thus not generalizing well to different datasets. Each of the asset prices was filtered before passing to the Latent Feature Extractor Unit.

\begin{figure}[h]
\includegraphics[width=\linewidth,height=8cm]{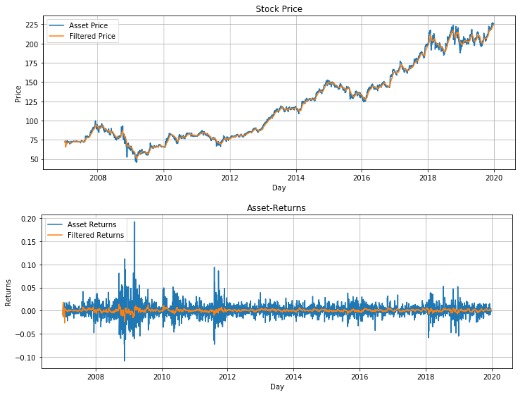}
\caption{Filtering Unit: Filtered Berkshire Hathaway Price and Returns Signal
}
\end{figure}

\subsection{Latent Feature Extractor Unit}

\textbf{Autoencoder Space}-
We build a time-series price encoder based on 3 encoder-decoder models- LSTM, CNN and DNN based autoencoders. We used a window size of 60 and the input consisted of price windows for each asset. To get better scalability of data, log returns were normalized by the min-max scalar. The data was divided into 80-20\% training-testing dataset. \par
The training process summary is as given in table 1, for both training and testing data. Number of epochs were decided on the convergence of mean-squared error of the autoencoder i.e $\textit{L}= \|X - X' \|^2$. \\

\textbf{ZoomSVD Space}-
Storage-Phase of ZoomSVD required dividing the input price data into blocks, and storing the initial asset-price matrix in compressed SVD form. The block parameter $b$ was taken as 60, but is usually decided on storage power of systems for large datasets. SVD of each block is stored by incremental-SVD. As our time-range query is passed by our RL agent, the SVD based state-space is computed by Partial and stitched-SVD. 

\begin{table}
\begin{center}
\begin{tabular}[h]{ |p{2cm}||p{1.5cm}|p{1.5cm}|p{1.5cm}|  }
 \hline
 \multicolumn{4}{|c|}{Autoencoder Training Summary} \\
 \hline
 & LSTM &CNN&DNN\\
 \hline
 Training set Mse   & 0.0610    &0.0567&   0.0593\\
 \hline
 Test set Mse&    0.0592 & 0.0531   &0.0554\\
 \hline
 Epochs &125 & 100&  100\\
 \hline
\end{tabular}
\end{center}
\caption{}
\label{table:ta}
\end{table}

\textbf{cRBM Space}-
We use a cRBM network stacked of 2 layers(1 visible layer, 1 hidden layer). Visible unit corresponding to the asset price window(60) and 30 neurons in the hidden layer for extracted feature space.

\begin{figure}[h]
\includegraphics[width=\linewidth ,height=8cm]{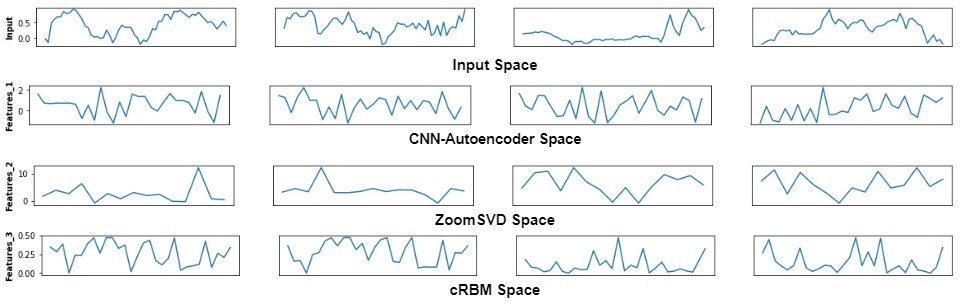}
\caption{Extracted Features from Input Price Window. Feature-1, 2, 3 corresponds to CNN-Autoencoder, ZoomSVD and cRBM extracted space, respectively
}
\end{figure}

\subsection{Training Step}
For each individual RL agent, number of episodes were decided on the convergence of percentage-distance from equal-weighted portfolio returns. A stable learning rate of 0.01 of adam optimizer was used, with 32 as batch size. The exploration probability is 20\% initially and decreases after every episode.

\subsection{Comparison of LFSS module based RL Agent with Baseline-RL agent}
After setting up LFSS module and integrating it to policy network, we backtested the results on given stock portfolio. Table 10 gives comparison summary of different feature extraction methods with one another and baseline-RL agent. Initial portfolio investment is 10000\$.  

As shown in Table 10, adding cRBM and CNN-Autoencoder based LFSS module to the agent leads to a significant increase in different performance metric, while DNN-Autoencoder based LFSS module gives slightly better performance than baseline-RL. ZoomSVD and LSTM-Autoencoder state space performs roughly similar to baseline-RL. Individual performance metric for each asset is described below- 
\begin{itemize}
\item Portfolio-Value(PV)- In terms of portfolio value over a span of 4 years on training-set an initial investment of 10000\$ reached roughly 39000\$ for test-set for CNN-Autoencoder and cRBM RL-agent, which performed significantly better than baseline RL-agent. The addition of autoencoder and RBM based LFSS module models the asset-market environment in a better way than raw price data for the RL-agent. Deep-learning based feature selection methods could model non-linear features in asset-price data as compared to linear space in ZoomSVD. Though cRBM RL agent did not give high returns in the initial years, it had a sharp increase in portfolio value in later years, performing much better than other methods during this period.
\item Sharpe-Ratio- Most of the models performed similarly in terms of volatility, with returns varying. There was no statistically significant difference in volatilities. This implies the LFSS module added is in not learning anything new to minimize risk in the model. This may be due to the simplicity of reward function used, and more work will be required to incorporate risk-adjusted reward to handle volatility, which fits in with DDPG algorithm.
\item Sortino Ratio- Addition of LFSS module, reduces downside deviation to some extent to the RL agent, but overall a more effective reward function can minimize downward deviation further.
\end{itemize}

\begin{figure}[h]
\includegraphics[width=\linewidth,height=6cm]{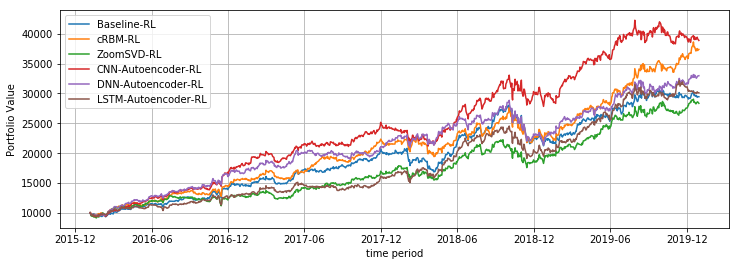}
\caption{Comparison of various RL agents
}
\end{figure}

\begin{table}
\begin{center}
\begin{tabular}[t]{ |p{1.4cm}||p{0.9cm}|p{0.9cm}|p{0.9cm}|p{0.9cm}|p{1cm}| }
 \hline
 \multicolumn{6}{|c|}{Performance Summary} \\
 \hline
 & Portfolio-Value(\$) & Annual Returns & Annual Volatility & Sharpe-Ratio & Sortino-Ratio  \\
 \hline
 Baseline RL   & 29440    & 0.486 &  0.254 &1.91&   2.12\\
 \hline
 DNN-Autoencoder RL &   32960  & 0.575   & 0.252 & 2.28 &   2.51\\
 \hline
 cRBM RL &   37368  & 0.684   &0.256 & 2.67 &   2.92\\
 \hline
 ZoomSVD RL &   28343  &0.458 & 0.257 & 1.78 &   2.03\\
 \hline
 CNN-Autoencoder RL &   38864  & 0.722   & 0.254 & 2.84 &   3.04\\
 \hline
 LSTM-Autoencoder RL &   30121  & 0.503   & 0.255 & 1.97 &   2.13\\
 \hline
\end{tabular}
\end{center}
\caption{Performance metric for different RL agents}
\label{table:ta}
\end{table}

\subsection{Performace Measure w.r.t to Benchmarks}
We used 5 non-RL based benchmarks for comparison of our agent with standard portfolio optimization techniques. The 5 benchmarks used were EW, S\&P 500, WMAMR, IMVK(Moderate and Aggressive Strategy) as explained in the last section. We compare these benchmarks to one other and to our RL-agent. The performance summary for all the methods is as described in table 3. 
\begin{itemize}
\item Portfolio-Value- In terms of portfolio value our RL agent and the baseline-RL performs much better than all the benchmarks used. This shows the superiority of deep RL based methods for the task of portfolio optimization. For the test set, equal-weighted portfolio gave better results than all our benchmarks. IMVK based strategies performed poorly on test set, mostly due to high change in weights leading to high transaction cost. If we consider the overall time period from 2007-2019 (fig 13), IMVK based strategies perform much better than equal-weighted portfolio till 2017, followed by a large drop in portfolio performance. 
\item Sharpe Ratio- Similar to high returns given by RL agents, deep RL based approach gave much better Sharpe Ratio than the benchmarks. Though there was high difference in volatility, the risk was compensated by the high returns.
\item Sortino Ratio- Similar to Sharpe ratio, RL agents performed much better to all other benchmarks. 
\end{itemize}

\begin{figure}[h]
\includegraphics[width=\linewidth,height=6cm]{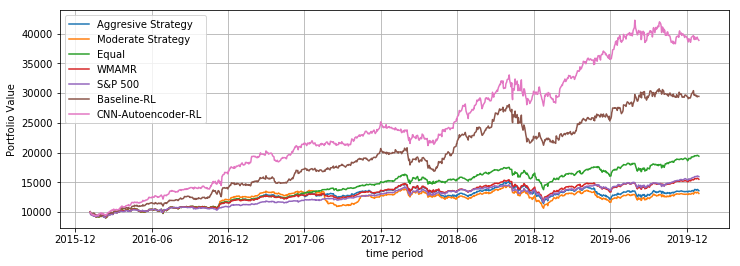}
\caption{Comparison of RL agent with various benchmarks
}
\end{figure}

\begin{figure}[h]
\includegraphics[width=\linewidth,height=6cm]{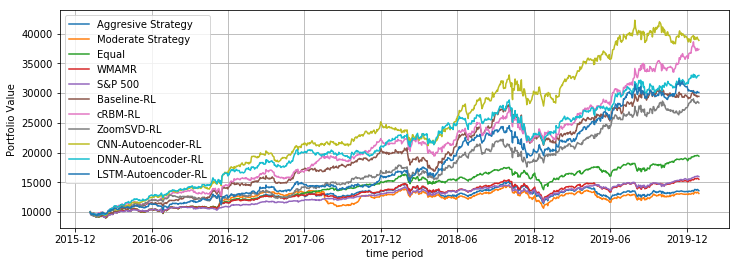}
\caption{Backtest Results for all methods proposed
}
\end{figure}

\begin{figure}[h]
\includegraphics[width=\linewidth,height=6cm]{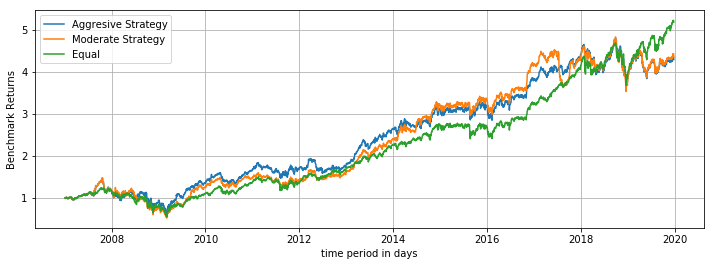}
\caption{Comparison of Aggressive and Moderate Strategy IMVK with Equal-Weighted Portfolio
}
\end{figure}

\begin{table*}
\begin{center}
\begin{tabular}[b]{ |p{1.1cm}||p{0.9cm}|p{1.27cm}|p{0.8cm}|p{1.1cm}|p{1.27cm}| p{1.27cm}|p{0.7cm}|p{1.1cm}|p{1.1cm}|p{0.8cm}| p{1cm}|}
 \hline
 \multicolumn{12}{|c|}{Performance Summary} \\
 \hline
 & Baseline-RL & DNN-Autoencoder RL & cRBM RL & ZoomSVD RL & CNN-Autoencoder RL & LSTM-Autoencoder RL & EW & Aggressive Strategy & Moderate Strategy & S\&P 500 & WMAMR \\
 \hline
 Portfolio Value(\$)   & 29440    &32960&   37368 & 28343 &  38864 & 30121   & 19456    &13604&   13144 & 15970 &   15524\\
 \hline
 Annual Returns &   0.486  & 0.575   &0.684&0.458&   0.722 & 0.503   & 0.235    & 0.0953 &   0.0785 & 0.149 &  0.152\\
 \hline
 Annual Volatility &   0.254  & 0.252   &0.256 &0.257 &   0.254 & 0.255   & 0.144    &0.162&   0.154&0.128 &  0.144\\
 \hline
 Sharpe Ratio &   1.91  &2.28   &2.67&1.78&   2.84 & 1.97   & 1.63    &0.59&   0.51 & 1.16 &   1.06\\
 \hline
 Sortino Ratio &   2.12  & 2.51   & 2.92 & 2.03 &   3.04 & 2.13   & 1.84    & 0.78 &   0.69 & 1.37 &   1.26\\
 \hline
\end{tabular}
\end{center}
\caption{Performance metric for all the proposed methods}
\label{table:ta}
\end{table*}   

\section{CONCLUSIONS}

In this paper, we propose a RL agent for portfolio optimization with an added LFSS Module to obtain a novel state space. We compared the performances of our RL agent with already existing approaches for a portfolio of 15 S\&P 500 stocks.  Our agent performed much better than baseline-RL agent and other benchmarks in terms of different metrics like portfolio value, Sharpe Ratio, Sortino Ratio, etc. The main catalyst for improvement in returns was due to the addition of LFSS module which consisted of a time series filtering unit and latent feature extractor unit, to get a compressed latent feature state space for the policy network. We further compared different feature space extracting methods like autoencoders, ZoomSVD and cRBM. Deep learning based techniques like CNN-Autoencoder and cRBM performed effectively in obtaining a state-space for our RL agent, that gave high returns, while minimizing risk. Additionally, we also introduced a new benchmark based on Kd-Index technical-indicator, that gave comparable performance to equal-weighted portfolio, and can be further used as a benchmark for more portfolio optimization problems. \par
The work done in this paper can provide a flexible modelling framework for various different state space with more sophisticated deep learning models and with the advancements in RL algorithms our RL agent can be further improved upon.

\section{Future Work}
Due to the flexible approach in state space modelling, there are many ways to improve upon the performance of the RL agent. To improve upon the quality of data through filtering layer, non-linear filters can be applied like EKF, particle filter, etc[32], which take into account the non-linearity in the data or use an adaptive filter[452] that at each time step uses most appropriate filter out of a set of filters. For extracting latent features, cRBM can be further extended to deep belief networks[43], which goes deeper than cRBM and is trained using a similar approach . Since RL is highly sensitive to the data used, our agent should be trained on more datasets, which are more volatile than American stock exchange data, as this will give a more robust performance. Additionally, we can add data augmentation module[26] to generate synthetic data, hence exposing the agent to more data. Text data based embedding can also be added to the state space[29], leveraging the use of NLP techniques. \par
In terms of deep RL algorithms, PPO was not used in this work. Performance of LFSS module with PPO algorithm can be further explored.
This framework can be integrated with appropriate risk-adjusted reward-function that minimizes the risk of our asset allocation.  

\section{Acknowledgement}
This work was supported by Professor Nitin Gupta and Professor Geetanjali Panda- Department of Mathematics. I would like to thank them for helpful insights on financial portfolio optimization, MDP and non-linear optimization, which
was essential in the progress of the project.



\end{document}